\documentclass{elsart}

 \usepackage{graphicx}

\usepackage{amssymb}

\begin{document}

\begin{frontmatter}



\title{On the heating of AGN magnetospheres}


\author{Osmanov Z.N.} 

\address{School of Physics, Free University of Tbilisi, 0183, Tbilisi, Georgia}
\address{E. Kharadze Georgian National Astrophysical Observatory, Abastumani 0301, Georgia}

\author{Mahajan S.}
\address{Institute for Fusion Studies, The University of Texas at
Austin, Austin, TX 78712, USA}

\begin{abstract}
The Langmuir-Landau-Centrifugal Drive (LLCD), that can effectively "convert" gravitational energy into particles, is explored as a driving mechanism responsible for the extreme thermal luminosity acquired by some active galactic nuclei (AGN). For this purpose we consider equations governing the process of heating of AGN magnetospheres. In particular, we examine the Fourier components of the momentum equation, the continuity equation and the Poisson equation in the linear approximation and estimate the growth rate of the centrifugally excited electrostatic waves and the increment of the Langmuir collapse. It is shown that the process of energy pumping is composed of three stages: in the first stage the energy is efficiently transferred from rotation to the electrostatic modes. In due course of time, the second regime - the Langmuir collapse - occurs, when energy pumping is even more efficient. This process is terminated by the Landau damping, when enormous energy is released in the form of heat. We show that the magnetospheres of the supermassive black holes with luminosities of the order of $10^{45-46}$erg/s can be heated up to $10^{6-10}$K.

\end{abstract}


\end{frontmatter}

\section{Introduction}

Rotating magnetospheres, with relatively strong 
magnetic fields, surrounding active galactic nuclei (AGN), are believed to be the regions where cosmic rays might originate {\bf \cite{rieger1,zev}}. This particles are highly relativistic and one of the open problems in modern astrophysics is how do they attain  such high energies.

Several mechanism have been suggested to explain acceleration of 
leptons or hadrons to ultra high energies. In particular, in the so-called Fermi class
(and its modifications) of processes \cite{fermi,bell1,bell2}, the acceleration occurs in 
relatively strong magnetic fields in the magnetospheres of astrophysical
objects. The Fermi processes, it turns out, are efficient only if the particles are already relativistic. For overall efficiency, therefore,  some kind of a pre-acceleration is desirable \cite{rieger1}. 

In this paper, however, we explore the Langmuir-Landau-Centrifugal Drive (LLCD) mechanism that makes use of the plasma collective phenomena manifested though the Langmuir wave that will serve as a conduit for transferring gravitational energy in the rotating magnetosphere of an AGN to particle kinetic /thermal energy. Although LLCD has been discussed and developed in several papers \cite{review,sgrA,screp1,zev}, we will summarise its salient qualitative features in the introduction.

Since in the AGN magnetospheres magnetic fields are of the order of $10^{1-4}$G, the charged particle flows will follow co-rotating magnetic field lines (frozen-in condition), experiencing very strong relativistic centrifugal 
force in the light cylinder (LC) zone (area where the linear velocity of rotation exactly equals
the speed of light) \cite{gold}. If the plasma number density is hight enough to provide the screening of longitudinal (along the magnetic field lines) electric field up to the LC area, the centrifugal mechanism becomes very efficient \cite{rieger2}. The strong relativistic differential centrifugal force leads to several consequences, for instance, the direct acceleration of electrons to very high energies (Lorentz factors of the order of $10^7$) \cite{rieger1,osm7}. Although the LLCD mechanism has been shown to accelerate particles to even higher energies, we in this paper emphasize its possible role in heating of the magnetosphere.

The first step in the LLCD process is the parametric excitation of Langmuir waves through charge separation caused by differential rotation of different charge species (electrons and protons ) endowed with different Lorentz factors. This process, examined for millisecond pulsars \cite{langm2} and AGN \cite{langm1}, proves to be very effective in pumping enormous energy (from rotation, which in turn is gravitationally driven) into what could be called centrifugally driven electrostatic waves. 
 
In the second stage of LLCD, the electrostatic waves Landau damp and transfer energy from 
centrifugally amplified modes to particle kinetic energy. The efficacy of this wave-particle energy transfer has been demonstrated for millisecond and newly born pulsars \cite{screp1,screp2}; it was shown that LLCD could energize particles to the tune of $10^{18}$eV. 

In the study of the role of LLCD in the AGN magnetospheres \cite{zev}, and SgrA$^{\star}$
(located in the centre of the Milky Way) \cite{sgrA}, we found that the parametrically amplified 
Langmuir waves, before undergoing standard Landau damp, go through another intermediate boosting via Langmuir collapse. One learns from the classical work by \cite{zakharov}, that the Langmuir turbulence in the non-linear regime creates unstably excited caverns or the low-density regions. As a result, the high
frequency pressure pulls the particles from the cavern, which in turn, provokes highly unstable 
collapse leading to fast amplification of the electrostatic field. The corresponding problem has been studied for three dimensional geometry by \cite{galeev} considering spectra of Langmuir turbulence. The termination by means of the Landau damping of the collapse process has been numerically studied by \cite{degtiarev}. This last stage is the most important one in energy pumping process, because the whole energy of Langmuir waves transfers to particle kinetic energies. Applying this process to AGN \cite{zev} and the central black hole of our galaxy \cite{sgrA}, it has been found that protons might achieve extremely high energies of the order of $ZeV$ (in AGNs) and $PeV$ (in SgrA$^{\star}$).

A principal consequence of  Langmuir collapse might be not only the particle acceleration in a particular direction but also the efficient heating of the sustaining plasma.  It is this heating that is posited here to be the cause of the observed intrinsic brightness (high temperatures ~$10^{6-9}$K) of some AGNs \cite{agn}.

The paper is organized   as follows: in Sec.2, the essential formulation of LLCD is recalled, in Sec. 3, the theoretical model is applied to AGN heating problem, and in Sec. 4, our principal results are summarized.

\section{The theoretical model of LLCD}

The plasma under consideration lies in the magnetosphere of a supermassive black hole of typical mass 
$M =10^{8-9}\times M_{\odot}$ ($M_{\odot}\approx 2\times 10^{33}$g is the solar mass), and corresponding angular velocity of rotation 
\begin{equation}
\label{rotat} \Omega\approx\frac{a c^3}{GM}\approx
10^{-3}\frac{a}{M_8}rad/s^2,
\end{equation}
where $c$ is the speed of light, 
$G\approx 6.67\times 10^{-8}$dyne-cm$^2$/g$^2$ is the gravitational constant and
$M_8\equiv M/(10^8M_{\odot})$ and $0<a\leq 1$ are respectively dimensionless mass and a
dimensionless parameter measuring the rate of rotation. 

If the plasma in the near vicinity of a supermassive black hole is anchored
by a magnetic field strong enough to maintain frozen-in condition, i.e, the  charged particles 
will follow the field lines. In our model, we assume that the AGN radiation energy density and magnetic field energy density are of the same orders of magnitude (approximate equipartition). Then, on the light cylinder surface, where the relativistic effects of rotation are the most efficient, the strength of magnetic field is given by \cite{osm7},
\begin{equation}
\label{mag} B\approx\sqrt{\frac{2L}{R_{lc}^2c}}\approx
870\times a\times L_{45}^{1/2}\times M_8 \;G,
\end{equation}
where $R_{lc}=c/\Omega$ is
the light cylinder radius and $L_{45}=L/\left(10^{45}erg/s\right)$ is the dimensionless 
luminosity. As it has been explained in detail by \cite{rieger1}, the charged particles co-rotate with the field lines until the Lorentz factor reaches its maximum value, $\gamma_{max}\sim\left(eB/(2m_p\Omega c)\right)^{2/3}\sim 1.2\times 10^7L^{1/3}_{45}$ (here $m_p$ is the proton's mass). Due accretion, the nearby area of AGN contains soft thermal photons which, by means of the IC scattering, might affect the process of acceleration. IC operates primarily on electrons while it is strongly suppressed for protons (by a factor ~$10^{-13}$) and does not limit the maximum energy protons may acquire. In the framework of our approach  the magnetic field lines almost up to the LC are supposed to be approximately straight \footnote{In the rotating magnetosphere this means that the particles in the rotating frame of reference move along straight trajectories. However, it has been shown that on the LC itself the field lines are swept back, lagging behind the rotation \cite{curve} but up to the mentioned zone the field lines are quasi straight.}

The original content of the LLCD mechanism, the  parametric generation of centrifugally driven Langmuir waves,  is contained in the set of linearized fluid equations (in Fourier space) coupled to the Poisson equation \cite{screp1,zev,screp2}:   
 \begin{equation}
\label{eul3} \frac{\partial p_{_{\beta}}}{\partial
t}+ik\upsilon_{_{\beta0}}p_{_{\beta}}=
\upsilon_{_{\beta0}}\Omega^2r_{_{\beta}}p_{_{\beta}}+\frac{e_{_{\beta}}}{m_{_{\beta}}}E,
\end{equation}
\begin{equation}
\label{cont1} \frac{\partial n_{_{\beta}}}{\partial
t}+ik\upsilon_{_{\beta0}}n_{_{\beta}}, +
ikn_{_{\beta0}}\upsilon_{_{\beta}}=0
\end{equation}
\begin{equation}
\label{pois1} ikE=4\pi\sum_{_{\beta}}n_{_{\beta0}}e_{_{\beta}},
\end{equation}
where Eq. (\ref{eul3}) is the equation of motion in the presence of a centrifugal force, Eq. (\ref{cont1}) represents the 
continuity equation, and Eq. (\ref{pois1}) is the Poisson equation for the electric field E. In the preceding equations,
 ${\beta}$ is the species index (either electrons or protons), $p_{_{\beta}}$ is the first order dimensionless momentum ($p_{_{\beta}}\rightarrow
p_{_{\beta}}/m_{_{\beta}}$), $k$ represents the wave number of the excited mode, $\upsilon_{_{\beta0}}(t) \approx
c\cos\left(\Omega t + \phi_{_{\beta}}\right)$ is the zeroth order
velocity and $r_{_{\beta}}(t) \approx
\frac{c}{\Omega}\sin\left(\Omega t + \phi_{_{\beta}}\right)$ is the
radial coordinate \cite{zev}, $e_{_{\beta}}$ is the particle's
charge and $n_{_{\beta}}$ and $n_{_{\beta0}}$ are, respectively the perturbed and
unperturbed Fourier components of the number density. 

The first term on the righthand side of Eq. (\ref{eul3}) represents the relativistic analogue of the centrifugal force. Since the centrifugal  force differentiates between different species (electrons and protons), the resulting charge separation excites the Langmuir instability.  

Notice that due to the time dependence of the centrifugal force, the preceding system cannot be "mode analyzed" by the standard techniques. However, we can extract almost as much information from the system as in a typical system, by the method developed in \cite{screp2}. The ansatz
\begin{equation}
\label{ansatz}
n_{\beta}=N_{\beta}e^{-\frac{iV_{\beta}k}{\Omega}\sin\left(\Omega t
+ \phi_{\beta}\right)},
\end{equation}
converts the aforementioned set of governing equations to a pair of coupled equations (time dependence is left only in $\chi$)
\begin{equation}
\label{ME1} \frac{d^2N_p}{dt^2}+{\omega_p}^2 N_p= -{\omega_p}^2 N_e
e^{i \chi},
\end{equation}
\begin{equation}
\label{ME2} \frac{d^2N_e}{dt^2}+{\omega_e}^2 N_e= -{\omega_e}^2 N_p
e^{-i \chi},
\end{equation}
where $\omega_{e,p}\equiv\sqrt{4\pi e^2n_{e,p}/m_{e,p}\gamma_{e,p}^3}$ is the relativistic 
plasma frequency of the corresponding specie, $\gamma_{e,p}$ is the Lorentz factor, $\chi = b\cos\left(\Omega t+\phi_{+}\right)$, $b =
\frac{2ck}{\Omega}\sin\phi_{-}$ and $2\phi_{\pm} = \phi_p\pm\phi_e$.

A quasi "dispersion relation" for the electrostatic wave is derived after a formal Fourier transform (For the detailed derivation please see the paper by \cite{screp2} of Eqs. (\ref{ME1},\ref{ME2}) 

\begin{equation}
\label{disp} \omega^2 -\omega_e^2 - \omega_p^2  J_0^2(b)= \omega_p^2
\sum_{\mu} J_{\mu}^{2}(b) \frac{\omega^2}{(\omega-\mu\Omega)^2},
\end{equation}
where $J_{\mu}(x)$ is the Bessel function. For resonant modes, $\omega = \mu\Omega+\Delta=\omega_r+\Delta$, $\Delta\ll \omega_r$, we find that the effective dispersion relation is contained in 
 \begin{equation}
 \label{disp1}
 \Delta^3=\frac{\omega_r {\omega_p}^2 {J_{\mu_{r}}(b)}^2}{2},
 \end{equation}
 that is readily solved to yield the instability growth rate  
\begin{equation}
 \label{grow1}
 \Gamma= \frac{\sqrt3}{2}\left (\frac{\omega_e {\omega_p}^2}{2}\right)^{\frac{1}{3}}
 {J_{\mu}(b)}^{\frac{2}{3}}.
\end{equation}
where $\mu = \omega_e/\Omega$. 

\begin{figure}
  \resizebox{\hsize}{!}{\includegraphics[angle=0]{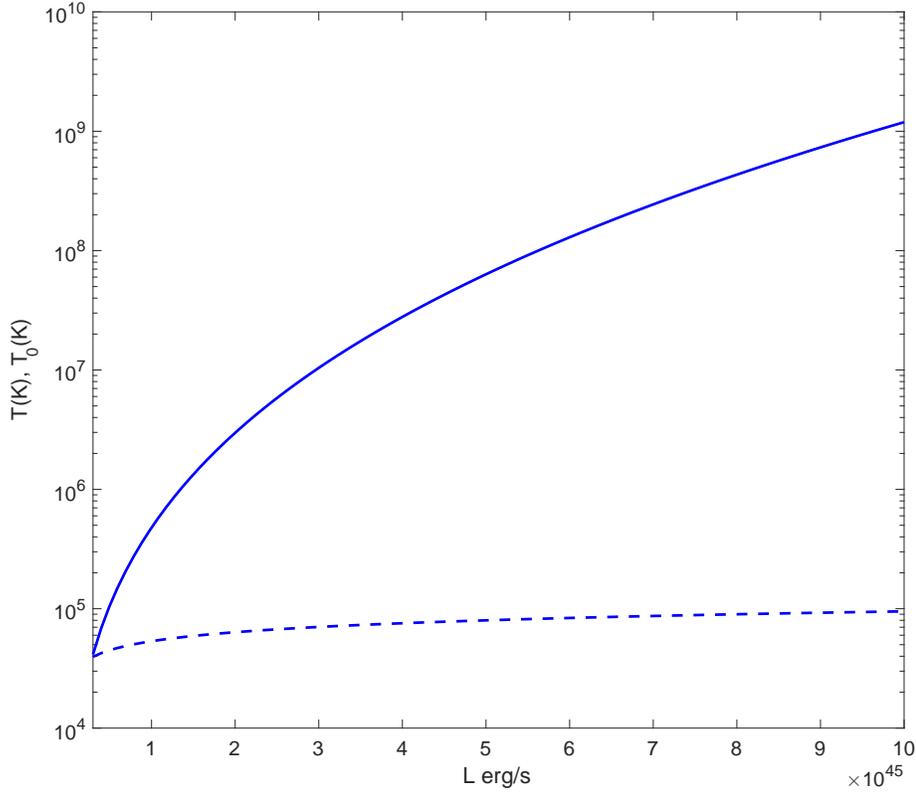}}
  \caption{Behaviour of $T$ (solid line) and $T_0$ (dashed line) with respect to the
  bolometric luminosity of AGN.
The set of parameters is: $a = 1$, $\epsilon = 0.1$, $\kappa = 0.5$, $M_8 = 1$, $\gamma_1 = 7\times 10^3$, $\gamma_2 =  10^4$ and $R_0\approx 0.01$pc.}\label{fig1}
\end{figure}

\section{Discusion}

Let us now examine the relevance of this instability as an agent for energy transfer in the AGN context.
Since this instability is driven by differential rotation between the electron and ion fluids, it falls in the general class of two stream instabilities. We now show that 
for the magnetospheric plasma, consisting of relativistic electrons and protons, the instability growth rates are "large" implying that  Langmuir wave generation  is very efficient. 

From Eq. (\ref{disp1}), one can straightforwardly show that for electrons with Lorentz factors $\gamma_1 = 7\times 10^3$, and protons having the same value $\gamma_2=10^4$, the timescale of energy pumping from rotation
to electrostatic waves, $\tau\sim 1/\Gamma$, varies in the range $600-2000$sec. On the other hand, the kinematic timescale (escape time-scale) of co-rotating particles (electrons and protons) equals $P/4=\pi/(2\Omega)\approx 1.5\times 10^4$sec. As it is evident, the instability timescale is much less than the kinematic timescale, indicating extremely high efficiency of centrifugally driven electrostatic waves.

These relatively high amplitude Langmuir waves will, now, induce high frequency
pressure pushing the particles out from the perturbed zone  \cite{zakharov}. In the resulting flow density 
areas (caverns), the penetrating waves will amplify pressure, which
in turn, augments the process of pulling out of the particles resulting in the Langmuir collapse.

We assume that the kinetic and potential energies of plasmons inside the caverns are of the same orders
of magnitude \cite{arcimovich}
\begin{equation}
\label{k} k^2\lambda_D^2\sim\frac{\mid\delta n\mid}{n_0},
\end{equation}
where $\lambda_D\equiv \sqrt{k_{B}T_0/(4\pi n_0e^2)}$ is the
Debye length-scale, $k_{B}\approx 1.38\times 10^{-16}$ erg K$^{-1}$
is the Boltzmann constant and $T_0$ is the temperature provided by the accretion process. Since the perturbation density is much less than the unperturbed density, $\delta n\ll n_0$, energy of plasmons is almost constant
\begin{equation}
\label{E2a} \int {d\bf r}\mid E\mid^2 = const.
\end{equation}

It is useful to recognize that  $k$ defines the system length-scale ($k\sim l^{-1}$).  Equation (\ref{E2a}), then, implies that the electrostatic energy density behaves as $E^2\propto l^{-q}$ where $q$ denotes the dimensionality of the process. Coupled with Eq.(\ref{k}), we find that the high frequency 
pressure, $P_{hf}\approx -E^2\delta n/(24\pi
k^2\lambda_D^2n_0)\propto E^2\propto l^{-q}$   \cite{arcimovich}, suppresses 
the thermal pressure, $P_{th} = k_BT_0\delta n\propto\delta n\propto k^2\propto l^{-2}$ only for three dimensional
geometry. Inside the magnetosphere, however, the effective geometry is one dimensional ($q=1$) since the particles are in the frozen-in condition and follow the field lines. In this region,  thus, the collapse
is impossible ($P_{hf}\sim l^{-1}$ and $P_{th}\sim l^{-2}$);  it can be realized only in the outer regions of the magnetospheres, for distances exceeding the LC radius.

Zakharov  showed that, for three dimensional geometry, $q=3$,
the driven electrostatic field and the corresponding length-scale of the cavern behave as
\cite{zakharov}
\begin{equation}
\label{E2} \mid E\mid\approx \mid E_0\mid\frac{t_0}{t_0-t}
\end{equation}
\begin{equation}
\label{l} l\approx l_0\left(\frac{t_0}{t_0-t}\right)^{-2/3},
\end{equation}
where $t_0$ is the time when the electric field fully collapses, and $E_0\approx 4\pi ne\Delta r\exp\left(\Gamma P/4\right)$ is the electrostatic field before the collapse starts (the initial field amplified by means of the electrostatic instability) and $\Delta r\approx
R_{lc}/(2\gamma_{_{max}})$ represents a length scale in the LC zone where the
process of energy pumping occurs \cite{zev} and we have taking into account that the escape time of particles equals $P/4$ \cite{MR}. It is clear from the preceding equations, that 
in due course of time, the length-scale of the cavern goes to zero, whereas the electrostatic field 
asymptotically increases. 

This process is terminated by means of Landau damping, when $l$ reaches the dissipation 
length-scale, $l_d\approx 2\pi\lambda_D$, \cite{arcimovich}. From Eqs. (\ref{E2},\ref{l}) one can
show that the electrostatic field will be boosted by the factor $\left(\Delta r/l_d\right)^{3/2}$. Correspondingly, a certain fraction, $\kappa$, of the electrical energy converts to heat, increasing the temperature of the ambient plasma,
\begin{equation}
\label{temp1} 2\kappa\pi R_{lc}l_dH\frac{E^2}{8\pi}\approx\frac{4\sigma}{c}T^4\pi R_{0}^2H,
\end{equation}
where (in the framework of the equipartition approach) it has been assumed that the energy pumped by means of the collapse nearby the LC zone in a thin 
layer, $l_d$ (being of the order of $(3-5)\times 10^3$cm), is uniformly distributed in a cylindrical area with typical radius, $R_{0}$; the energy is emitted away as a black body radiation. Here we assume that the typical heated radius  is of the order of $0.01$pc (an approximate outer radius of the accretion disc for the given mass of the BH \cite{carroll}, and $H$ represents the height of the corresponding cylinder. From Eq. (\ref{temp1}) one can, straightforwardly, obtain the increased temperature of the AGN ambient
\begin{equation}
\label{temp} T\approx\left(\frac{c\kappa l_d R_{lc}E^2}{16\pi\sigma R_{0}^2}\right)^{1/4}.
\end{equation}
The proposed heating mechanism is efficient, if the corresponding temperature exceeds that of the 
accretion disk temperature, which is estimated as \cite{carroll}
\begin{equation}
\label{T0} T_0\approx\left(\frac{3GM\dot{M}}{8\pi\sigma R^3}\right)^{1/4}\left(\frac{R}{R_{lc}}\right)^{3/4}
\left(1-\sqrt{\frac{R}{R_{lc}}}\right)^{1/4},
\end{equation}
where $R\equiv 2GM/c^2$ is the Schwarzschild radius of the supermassive black hole, $\dot{M}=L/(\epsilon c^2)$ is the accretion rate and $\epsilon\leq 1$ describes efficiency of the accretion process.

In Fig. 1 we show the dependance of $T_0$ and $T$ on the bolometric luminosity of AGN. The set of parameters is: $a = 0.1$, $\epsilon = 0.1$, $\kappa = 0.5$, $M_8 = 1$, $\gamma_1 = 7\times 10^3$, $\gamma_2 =  10^4$, and $R_0 \approx 0.01$pc.
As is evident from the plots, the centrifugally driven heating mechanism becomes efficient for highly luminous AGN. In particular, for AGN with $L>>3\times 10^{44}$erg/s the temperature becomes much higher than the initial value ($T_0\sim 10^5$K) acquired in the accretion processes. On the other hand, one can straightforwardly check that the heating process is not sensitive with the rotation rate, which is a direct result of the power $1/4$ (see Eq. (\ref{temp})). Here we assumed the equipartition distribution of energy: half of the pumped energy goes to acceleration and half - to the heating process.

The heating mechanism considered in the present work is so efficient that it can heat the system to blackbody radiation temperatures~$10^{9}$K, which have been observed in the $X$-ray corona of AGNs \cite{agn}. 

It is worth noting that during the acceleration the particles might lose energy by means of the several cooling mechanisms. This problem is considered in detail by \cite{zev}.The synchrotron process almost from the very beginning of motion becomes insignificant because the corresponding radiation is so efficient that the particles very soon transit to the ground Landau level, follow the magnetic field lines and do not emit any more in the synchrotron regime.

The curvature radiation does not impose any significant constraints as well. In particular, on the last stage of LLCD the particles achieve such high energies that the magnetic field becomes dependent on plasma energy density, the particles move along straight trajectories and as a result the curvature radiation is terminated.

The inverse Compton (IC) scattering for the considered energies occurs in the Klein-Nishina regime, when the corresponding timescale is a continuously increasing function of the proton energy \cite{blumenthal}, therefore this mechanism does not limit maximum achievable energies. 

Another mechanism, which potentially might limit the process of proton acceleration is the photo-pion energy losses. The corresponding timescale is a continuously decreasing function of energy and for the extremely high values $10^{19-20}$eV is of the order of $10^7$sec. On the other hand, the timescale of energy pumping into Langmuir waves is by several orders of magnitude less than the aforementioned value. The collapse in turn, since characterised by blowing up (see Eq. (\ref{E2}-\ref{l})) is even more efficient than any other mechanisms. Therefore, the photo-pion cooling does not impose any constraints on proton acceleration and consequently on the heating process.

\section{Summary}

By examining the LLCD mechanism we have considered the system of equations composed 
by the Euler equation, continuity equation and the Poisson equation respectively. It has been shown 
that by means of the centrifugal force, the Langmuir waves parametrically amplify. As a result, 
the excited electrostatic waves efficiently pump energy from rotation.

Amplification of electrostatic field on the next stage is provided by means of the Langmuir collapse, which terminates on the Debye length-scales, resulting in the heating of the surrounding medium. It has been shown that for luminosities greater than $2\times 10^{45}$erg/s the mentioned process can provide temperatures 
in the following interval $10^{6-9}$K.

\bibliographystyle{cas-model2-names}



\end{document}